# Hydrogen-regulated chiral nanoplasmonics

*Xiaoyang Duan[1,2†], Simon Kamin[1,2†], Florian Sterl[3], Harald Giessen[3], and Na Liu[1,2]\**

[1]Max Planck Institute for Intelligent Systems, Heisenbergstrasse 3, 70569 Stuttgart, Germany
[2]Kirchhoff Institute for Physics, University of Heidelberg, Im Neuenheimer Feld 227, 69120 Heidelberg, Germany
[3]4th Physics Institute and Research Center SCoPE, University of Stuttgart, 70550 Stuttgart, Germany
†These authors contributed equally to this work.
\*E-mail: laura.liu@is.mpg.de

**ABSTRACT**:

Chirality is a highly important topic in modern chemistry, given the dramatically different pharmacological effects that enantiomers can have on the body. Chirality of natural molecules can be controlled by reconfiguration of molecular structures through external stimuli. Despite the rapid progress in plasmonics, active regulation of plasmonic chirality, particularly in the visible spectral range, still faces significant challenges. In this Letter, we demonstrate a new class of hybrid plasmonic metamolecules composed of magnesium and gold nanoparticles. The plasmonic chirality from such plasmonic metamolecules can be dynamically controlled by hydrogen in real time without introducing macroscopic structural reconfiguration. We experimentally investigate the switching dynamics of the hydrogen-regulated chiroptical response in the visible spectral range using circular dichroism spectroscopy. In addition, energy dispersive X-ray spectroscopy is used to examine the morphology changes of the magnesium particles through hydrogenation and dehydrogenation processes. Our study can enable plasmonic chiral platforms for a variety of gas detection schemes by exploiting the high sensitivity of circular dichroism spectroscopy.

Controlling molecular chirality is of great importance in stereochemistry, as enantiomers of a given compound may give rise to strikingly different biological activity.[1–7] Chirality of natural molecules can be manipulated by reconfiguring molecular structures through light[1–4], electric field[5], and thermal stimuli[6,7]. However, such chirality regulation is not prominent, in that the circular dichroism (CD) response of natural molecules is very weak. Recent advances in plasmonics renders the realization of artificial chiral metamolecules possible.[8–19] These plasmonic metamolecules consist of metallic structures that are arranged in chiral configurations. They can exhibit pronounced CD that is several orders of magnitude stronger than that of natural molecules. Nevertheless, reconfiguring plasmonic metamolecules, which are restrained on substrates, faces significant challenges. In 2012, S. Zhang *et al.* have proposed a seminal concept of chirality switching without structural reconfiguration in the terahertz range.[20,21] The underlying idea is to break the mirror symmetry of the structure through the integrated silicon component upon photoexcitation. Later, a thermally-controlled plasmonic chiral system in the mid-infrared range has been demonstrated through integration of $Ge_3Sb_2Te_6$.[22,23] As a result, utilization of active materials, in particular phase-transition materials, brings about an elegant solution to regulating plasmonic chirality without structural reconfiguration. This is in stark contrast to the case of natural molecules.

Among a variety of phase-transition materials, magnesium (Mg) possesses a unique role due to its multifunctional character.[24–29] First, Mg can undergo phase-transition from metal to dielectric to form magnesium hydride ($MgH_2$) upon hydrogen loading. $MgH_2$ can contain up to 7.6 wt % of hydrogen, featuring a remarkable capacity that exceeds all known reversible metal hydrides. Together with its abundance and low-cost aspects, Mg bears great potential for hydrogen-related applications. Second, different from other phase-transition metals for hydrogen storage, such as yttrium (Y)[30] and palladium (Pd)[31–35], which yield poor plasmonic properties, Mg constitutes a promising candidate for high-frequency plasmonics due to its excellent extinction efficiencies in the UV and blue visible wavelength range[27].

In this Letter, we demonstrate a hydrogen-regulated chiroptical response in hybrid plasmonic metamolecules in the visible spectral range. Here, Mg works not only as active material for hydrogen uptake but also as plasmonic material for resonant coupling with the satellite gold (Au) particles. Each metamolecule consists of coupled hybrid components that are arranged in a prescribed chiral geometry to generate a chiroptical response. Au particles are particularly employed here to assist Mg particles for achieving sharp and pronounced CD spectra. Such chiroptical response can be switched on/off by dynamically unloading/loading hydrogen in real time. In addition, energy dispersive X-ray spectroscopy (EDX) reveals the morphology changes of the Mg particles that result from hydrogenation and dehydrogenation processes. With integration of appropriate active materials, our design scheme can be generalized to create sensitive chiral platforms for a variety of gas detection, given the high sensitivity of CD spectroscopy. This could provide a powerful addition to the conventional sensing paradigm.

Figure 1a shows the schematic of the hybrid plasmonic metamolecules. Four Mg particles are surrounded by four satellite Au particles. These closely-spaced particles reside on a glass substrate and form a gammadion-like hybrid superstructure.[10,22] To facilitate hydrogen loading/unloading in Mg at room temperature, titanium (Ti, 5 nm) and Pd (10 nm) are capped on top of the Mg particles. Pd serves as catalyst to dissociate molecular hydrogen into atomic hydrogen.[25,26] Ti is used as spacer to prevent Mg and Pd from alloying as well as to release the mechanical stress resulting from different volume expansions of Mg (32%) and Pd (11%) upon hydrogen uptake.[28,29] The structural parameters of the Mg and Au particles are designed such that they can be *resonantly* coupled and consequently the chiral plasmonic metamolecule can yield different absorbance in response to left- and right-handed circularly polarized light, *i.e.*, CD. The different refractive indices of the substrate and air as superstrate are the symmetry-breaking factors that lead to the required CD response.

The samples were fabricated with a double electron-beam lithography process.[31] An exemplary scanning electron microscopy (SEM) image of a left-handed sample is presented in Fig. 1b, where the Mg and Au particles can be clearly distinguished due to the high contrast between these two materials. The Mg particles show more grainy features when compared to the Au particles. A tilted-view of one enlarged structure is presented in the inset.

The working principle of hydrogen regulation to the chiroptical response is illustrated in Figs. 2a and 2b. Initially, the left-handed plasmonic metamolecules exhibit a characteristic peak-to-dip line shape in the CD spectrum, which is simulated using the commercial software COMSOL (see the red line in Fig. 2b). This stems from the resonant coupling between the collective plasmons excited in the eight plasmonic particles within one hybrid metamolecule. As shown by the simulated field distribution in Fig. 2a, the electromagnetic near-fields are strongly localized between the Mg and Au particles. Upon hydrogen loading, Pd catalyzes the dissociation of hydrogen molecules into hydrogen atoms, which then diffuse through the Ti spacer into the Mg particle.[25,26] Subsequently, the center four Mg particles are hydrogenated into $MgH_2$, undergoing a phase-transition from metal to dielectric.[29] As a result, the plasmonic metamolecule becomes achiral, giving rise to a featureless CD spectrum (see the blue line in Fig. 2b). The corresponding field distribution reveals that only the four far-spaced Au particles are excited in this case.

The dynamics of hydrogen-regulation to the chiroptical response is investigated by placing the respective samples in a custom gas cell and measuring the CD spectra with a Jasco-1500 CD spectrometer in real time. All the measurements are carried out at room temperature. Studies of reaction kinetics in dependence on temperature are in progress. As shown in Fig. 2c, before hydrogen loading, the left-handed sample exhibits a distinct bisignate spectral profile (see the thick red line), which agrees well with the simulated spectrum in Fig. 2b. Subsequently, the sample is exposed to 0.25 vol. % hydrogen (in nitrogen carrier gas). The CD spectra are continuously recorded with an interval of 10 min. The CD strength successively decreases over time until the bisignate spectral profile vanishes eventually (see the thick blue line in Fig. 2c). The whole hydrogenation process takes approximately 100 min to complete. Hence, the chiroptical response of the plasmonic metamolecules can be dynamically switched off by hydrogen.

Previously, it has been demonstrated that hydrogen can be unloaded from Mg-based films in the presence of oxygen[24–26]. Such oxidative dehydrogenation has also recently been observed in Mg nanoparticles that were exposed to oxygen in nitrogen carrier gas[29]. To provide deeper insight into dehydrogenation of Mg nanoparticles under realistic environment conditions, the hydrogenated left-handed sample is exposed to ambient air at room temperature. For a direct comparison, the CD spectrum presented by the thick-blue line in Fig. 2c is replotted in Fig. 2d, as it is the starting curve for the dehydrogenation process. It is observed that hydrogen unloading from the Mg particles proceeds quite slowly. The measured intermediate CD spectra can be found in Fig. S1. The chiroptical response of the plasmonic metamolecules can be recovered after 27 hours in ambient air. The recovered CD spectrum (see the red line in Fig. 2d) exhibits slight alterations when compared to that in Fig. 2c (see the thick red line). This is due to morphology changes of the Mg particles through hydrogenation and dehydrogenation processes, which will be discussed in detail later.

Subsequently, the left-handed sample is exposed to hydrogen for the second cycle. It turns out that 0.25 vol. % hydrogen cannot introduce rapid CD changes any longer. A higher hydrogen concentration of 0.5 vol. % is then utilized. The chiroptical response is gradually switched off after 120 min (see Fig. S1), longer than the operation time in the first cycle, where hydrogen at only half of that concentration had been applied. This reflects a substantial influence of the morphology changes of the Mg particles on the hydrogen loading efficiency. The sample is then placed in ambient air for dehydrogenation again. The recovered chiroptical response shows degradation compared to that in the first cycle. In the third cycle, a higher hydrogen concentration of 2.0 vol. % is required to switch off the chiroptical response after approximately 60 min (see Fig. S1). The experimental results from these three cycles unambiguously indicate a close relationship between the hydrogen concentration and the switching dynamics of the plasmonic metamolecules, which warrants further investigation.

To provide deeper insight into the hydrogen loading efficiency change, the left-handed sample exposed to ambient air for one week after three cycles of hydrogen loading/unloading has been examined by EDX. A fresh left-handed sample has also been analyzed together for a direct comparison. EDX mapping offers a detailed illustration of the present elements and their localization in the samples. The EDX results for the fresh and used samples are shown in Figs. 3a and 3b, respectively. It is evident that Au, Ti, and Pd do not really show dissimilarity between the two samples, whereas Mg exhibits remarkable transformations. More specifically, in the fresh sample, the EDX signal of Mg from the centered four particles is very distinct and localized. On the contrary, in the used sample, the EDX signal of Mg becomes largely spread out.

This is likely due to cracking of the Mg particles, resulting from significant hydrogen-induced expansion and contraction (32%).[29] Consequently, the Mg particles become brittle and prone to be fragmented over the surroundings. Meanwhile, chemical compounds containing considerable carbon are found to accumulate around the central areas, where the four Mg particles are located originally. It is known that Mg, as a chemically active material, can be involved in several gaseous reactions. In ambient air, Mg can form magnesium oxides with oxygen, magnesium hydroxycarbonate with carbon dioxide, and possibly becomes hydroxylated due to formation of brucite ($Mg(OH)_2$) in a humid environment.[36,37] These Mg-based chemical compounds may first form on the particle surfaces in ambient air. Along with the cracking, the inner part of the particles is also gradually decomposed into Mg-based chemical compounds. As a result, the chiroptical response of the plasmonic metamolecules is eventually switched off as depicted by the green line in Fig. 2c. This CD spectrum is measured after the left-handed sample has been cycled through hydrogen loading/unloading three times, followed by exposure to ambient air for one week. It is noteworthy that if the hydrogenated Mg particles were exposed to oxygen in nitrogen carrier gas, due to the formation of magnesium oxide, certain morphological changes would still exist, resulting from repetitive structural expansions and contractions during hydrogenation and dehydrogenation processes.

To further investigate the relationship between the hydrogen concentration and the switching dynamics of the plasmonic metamolecules, CD spectra of two identical samples in right-handedness are measured using a time-scan function of the CD spectrometer in real time. A representative CD spectrum is shown by the red line in the inset of Fig. 4. It displays a dip-to-peak line shape, which is mirrored to the CD spectrum of the left-handed sample in Fig. 2b. The two right-handed samples are exposed to 0.3 vol. % and 3.0 vol. % hydrogen, and the corresponding CD spectra are recorded at a fixed wavelength of 705 nm, respectively. The results are shown in Fig. 4, where several particular features are observed. First, at a low concentration of 0.3 vol. %, a short accumulation time (~ 2 min) elapsed before the system starts to respond to hydrogen, whereas at a high concentration of 3.0 vol. %, the response to hydrogen loading is almost instant, giving rise to an abrupt CD change. Second, in stark contrast to Pd hydrogenation, which is concentration dependent, Mg hydrogenation at different concentrations can reach the same ending hydride state as long as the hydrogen loading time is sufficient. The chiroptical response is switched off after 25 min and 70 min at hydrogen concentrations of 3.0 vol. % and 0.3 vol. %, respectively. The featureless CD spectrum after hydrogenation is presented by the blue line in the inset of Fig. 4.

To corroborate our results, several control experiments have been performed. The CD response of an achiral sample, where the four Au particles are positioned along the center gaps between the four Mg particles, is examined first. As displayed in Fig. 5a, hydrogen loading does not introduce significant influence on the achiral sample. The negligible CD features stem from the alignment imperfection between the Au and Mg particles during sample fabrication. Furthermore, as shown in Fig. 5b, the four satellite Au particles and the four Mg particles alone exhibit unstructured and noisy CD spectra. They do not respond to hydrogen loading. Finally, a chiral sample composed of pure Mg particles is investigated and the results can be found in Fig. S2. In principle, it can also serve for generating hydrogen-regulated chiroptical response, however, giving a much weaker CD magnitude and therefore lower sensitivity when compared to the Au particle assisted scheme (see Fig. 2).

In conclusion, we have demonstrated hybrid plasmonic metamolecules, which allow for a hydrogen-regulated chiroptical response in the visible range. To create hydrogen-regulated plasmonic chirality, Mg might be one of the few metals which can respond to hydrogen and simultaneously offer resonant coupling with Au particles in the visible range. The switching dynamics has been investigated using CD spectroscopy in real time. The chiroptical response of the plasmonic metamolecules can be switched on/off through hydrogen unloading/loading in cycles. In addition, EDX mapping has been carried out to investigate the morphology changes of the Mg particles placed in ambient air after hydrogen loading/unloading in cycles. Our results suggest that due to its active chemical reactivity, Mg could be used as plasmonic sensors for humidity, carbon dioxide, and hydrogen detection, if specific surface protection is taken into consideration. Also, different from Pd and Y, which release the stored hydrogen as soon as hydrogen in the environment drops, Mg can preserve its stored hydrogen, as long as oxygen is absent at room temperature. Therefore, Mg can be retained at any hydrogenated state between fully metallic and fully dielectric. In this regard, our chiral plasmonic platform could enable the realization of dynamically tunable circular polarizers and polarization modulators regulated by hydrogen.

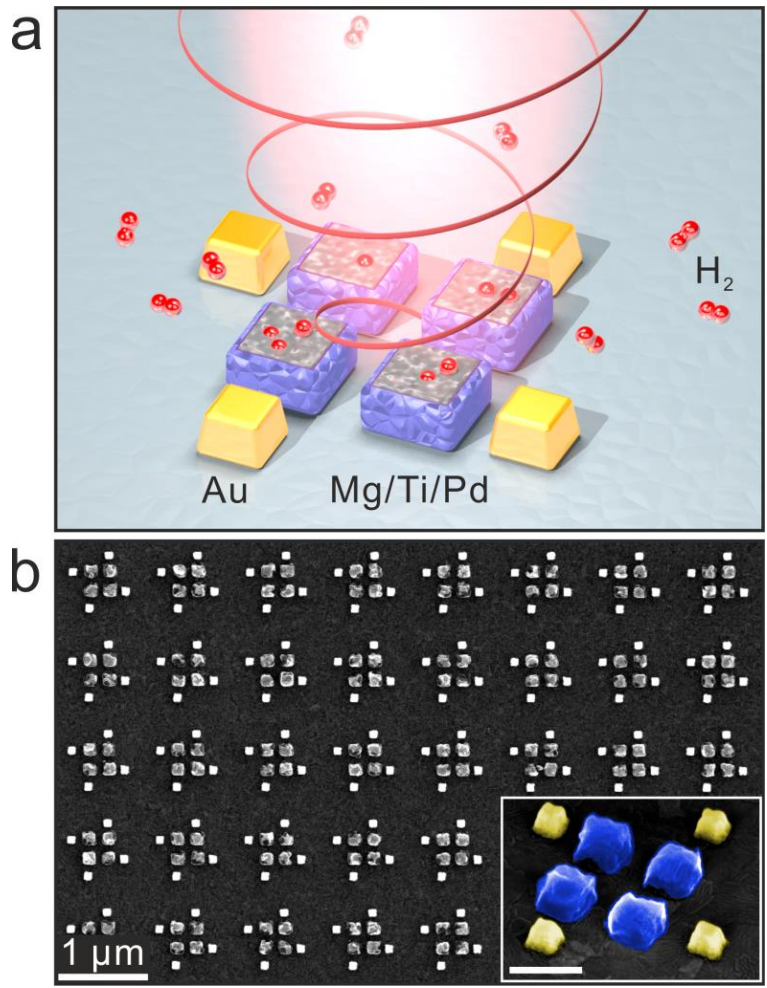

**Figure 1.** (a) Schematic of a hybrid plasmonic metamolecule on a glass substrate. Four Mg particles (160 nm × 160 nm × 75 nm) are surrounded by four Au satellite particles (100 nm × 100 nm × 80 nm). The Mg particles are capped with a 5 nm Ti spacer and a 10 nm catalytic Pd layer. The gaps between the Mg particles are 80 nm. The gaps between the Mg and Au particles are 60 nm. Circularly polarized light is incident perpendicularly onto the structure. Hydrogen is represented by the red spheres. (b) Overview SEM image of the left-handed hybrid plasmonic structures. The period along both directions is 1 μm. The scale bar in the tilted SEM image is 200 nm.

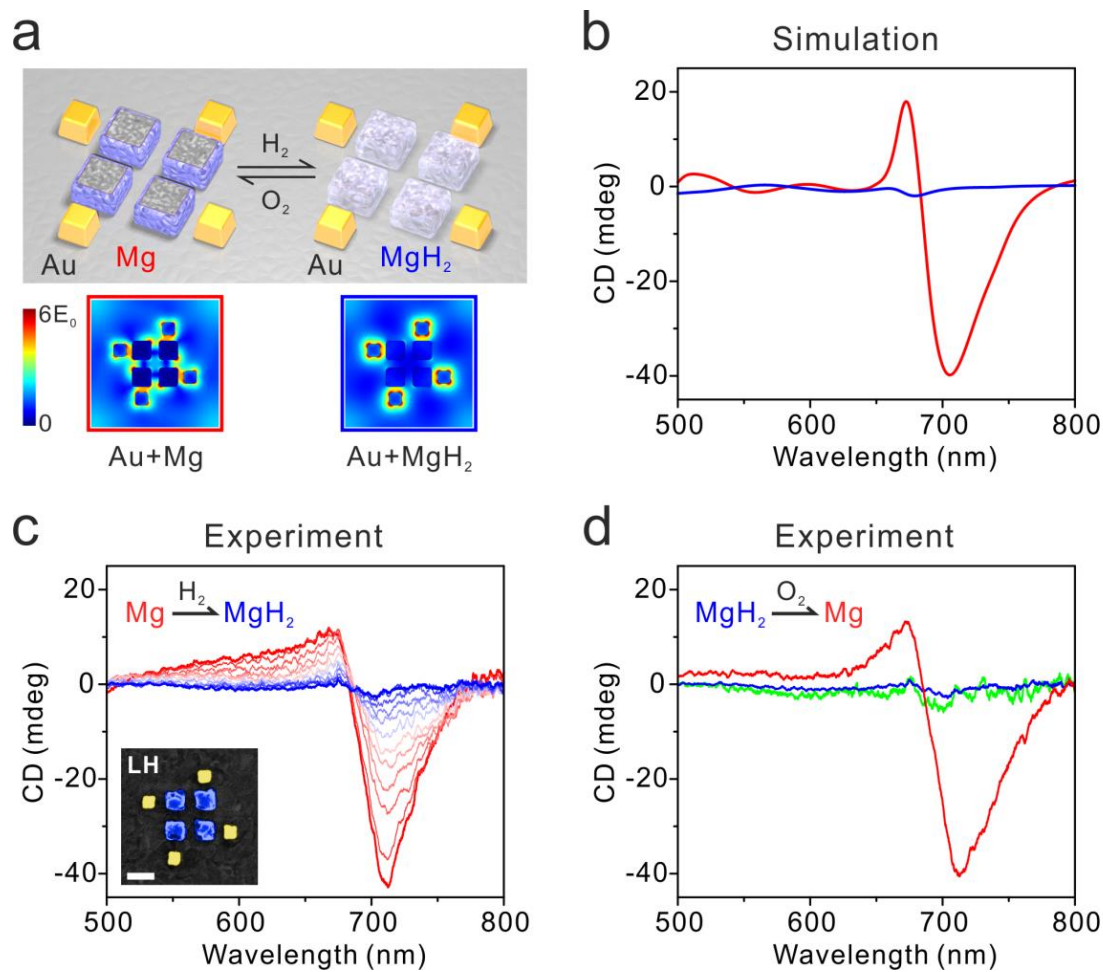

**Figure 2.** (a) Working principle of hydrogen regulation to the chiroptical response. Simulated field distributions of the hybrid structure before and after hydrogen loading. (b) Simulated CD spectra of the hybrid structure before (red) and after (blue) hydrogen loading. (c) Evolution of the measured CD spectra of the left-handed sample upon hydrogen loading as a function of time. The CD spectra are continuously recorded with an interval of 10 min. The distinct CD response (thick red) is switched off (see the thick blue line) by 0.25 vol. % hydrogen after 100 min. The scale bar in the SEM image is 200 nm. (d) Dehydrogenation of the left-handed sample in ambient air. The featureless CD spectrum (blue) can be recovered after 27 hours (see the red line). The CD spectrum in green is measured after the sample has been cycled through hydrogen loading/unloading three times, followed by being exposed to ambient air for one week.

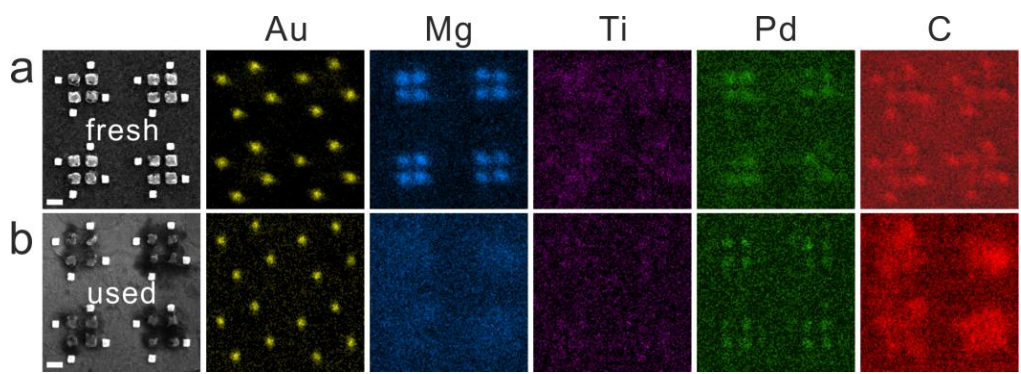

**Figure 3.** EDX mapping of (a) fresh and (b) used left-handed samples. The left-handed sample has been cycled through hydrogen loading/unloading three times, followed by exposure to ambient air for one week. The scale bars in the SEM images are 200 nm.

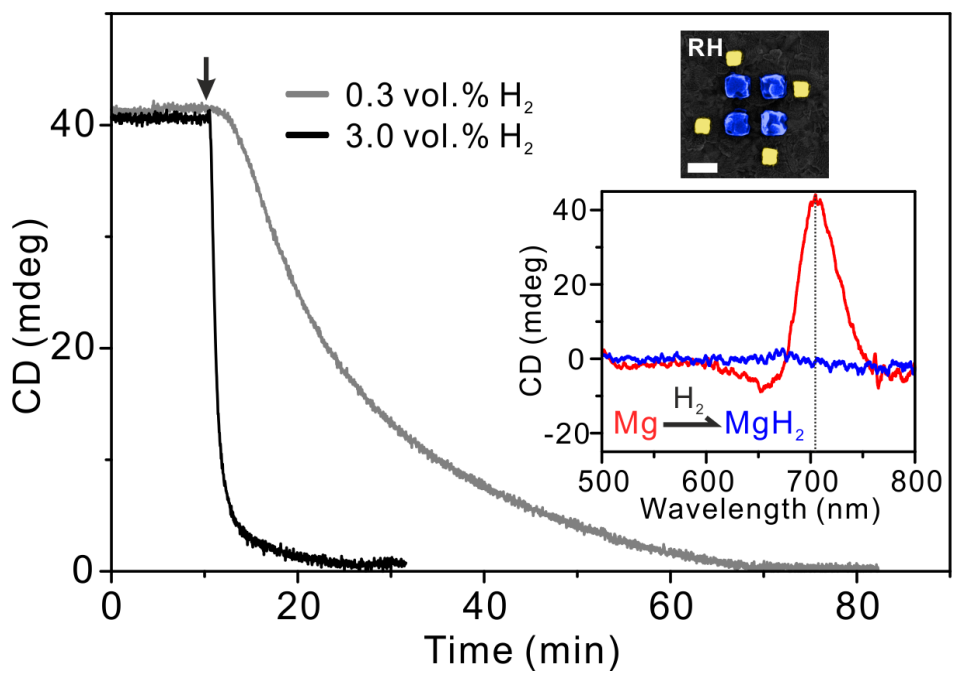

**Figure 4.** Switching dynamics of the hybrid plasmonic metamolecules at different hydrogen concentrations. The CD spectra of two identical right-handed samples are recorded using the real time-scan function of the CD spectrometer at a fixed wavelength of 705 nm. The measured CD spectra before and after hydrogen loading are depicted by the red and blue lines in the inset, respectively. A representative SEM image of the right-handed structure is also shown as inset. The scale bar in the SEM image is 200 nm.

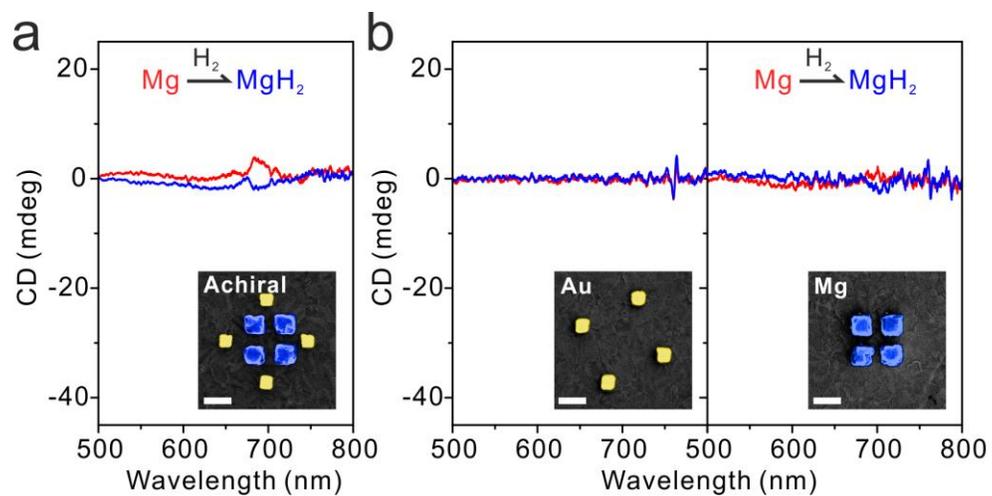

**Figure 5.** Control samples before (red) and after (blue) hydrogen loading. The applied hydrogen concentration is 3.0 %. (a) Achiral sample. (b) Satellite Au particles and centered Mg particles. All these samples do not show substantial chiral response to hydrogen loading. The scale bars in the SEM images are 200 nm.

**Supporting Information**. Sample fabrication, theoretical calculations, and additional experimental data. This material is available free of charge via the Internet at http://pubs.acs.org.

AUTHOR INFORMATION

**Corresponding Author**

*E-mail: laura.liu@is.mpg.de

**Notes**

The authors declare no competing financial interest.

**Acknowledgements**

We gratefully acknowledge the generous support by the Max-Planck Institute for Solid State Research for the usage of clean room facilities. We thank B. Fenk for acquiring the EDX data. This project was supported by the Sofja Kovalevskaja grant from the Alexander von Humboldt-Foundation, the Marie Curie CIG grant, and the European Research Council (ERC *Dynamic Nano,* ERC *ComplexPlas*) grants. We also acknowledge support from DFG, BMBF, MWK Baden-Württemberg, Zeiss-Stiftung, and BW-Stiftung (Spitzenforschung II, IFOG).

**References**

(1) Huck, N. P. M.; Jager, W. F.; de Lange, B.; Feringa, B. L. *Science* **1996**, *273*, 1686–1688.
(2) Koumura, N.; Zijlstra, R. W. J.; van Delden, R. A.; Harada, N.; Feringa, B. L. *Nature* **1999**, *401*, 152–155.
(3) de Jong, J. J. D.; Lucas, L. N.; Kellogg, R. M.; van Esch, J. H.; Feringa, B. L. *Science* **2004**, *304*, 278–281.
(4) Ikeda, K.; Liu, W.; Shen, Y. R.; Uekusa, H.; Ohashi, Y.; Koshihara, S. Y. *J. Chem. Phys.* **2005**, *122*, 141103.
(5) Amaranatha Reddy, R.; Schröder, M. W.; Bodyagin, M.; Kresse, H.; Diele, S.; Pelzl, G.; Weissflog, W. *Angew. Chem. Int. Edit.* **2005**, *44*, 774–778.
(6) Murata, K.; Aoki, M.; Suzuki, T.; Harada, T.; Kawabata, H.; Komori, T.; Ohseto, F.; Ueda, K.; Shinkai, S. *J. Am. Chem. Soc.* **1994**, *116*, 6664–6676.
(7) Hasegawa, T.; Morino, K.; Tanaka, Y.; Katagiri, H.; Furusho, Y.; Yashima, E. *Macromolecules* **2006**, *39*, 482–488.
(8) Rogacheva, A. V.; Fedotov, V. A.; Schwanecke, A. S.; Zheludev, N. I. *Phys. Rev. Lett.* **2006**, *97*, 177401.
(9) Gansel, J. K.; Thiel, M.; Rill, M. S.; Decker, M.; Bade, K.; Saile, V.; von Freymann, G.; Linden, S.; Wegener, M. *Science* **2009**, *325*, 1513–1515.
(10) Hendry, E.; Carpy, T.; Johnston, J.; Popland, M.; Mikhaylovskiy, R. V.; Lapthorn, A. J.; Kelly, S. M.; Barron, L. D.; Gadegaard, N.; Kadodwala, M. *Nat. Nanotechnol.* **2010**, *5*, 783–787.
(11) Kuzyk, A.; Schreiber, R.; Fan, Z.; Pardatscher, G.; Roller, E.-M.; Högele, A.; Simmel, F. C.; Govorov, A. O.; Liedl, T. *Nature* **2012**, *483*, 311–314.
(12) Shen, X.; Asenjo-Garcia, A.; Liu, Q.; Jiang, Q.; de Abajo, F. J. G.; Liu, N.; Ding, B. *Nano Lett.* **2013**, *13*, 2128–2133.
(13) Shen, X.; Zhan, P; Kuzyk, A.; Liu, Q.; Asenjo-Garcia, A.; Zhang, H.; de Abajo, F. J. G.; Govorov, A.; Ding, B.; Liu, N. *Nanoscale* **2014**, *6*, 2077–2081.

(14) Kuzyk, A.; Schreiber, R.; Zhang, H.; Govorov, A. O.; Liedl, T.; Liu, N. *Nat. Mater.* **2014**, *13*, 862–866.
(15) Zhou, C.; Duan, X.; Liu, N. *Nat. Commun.* **2015**, *6*, 8102.
(16) Hentschel, M.; Schäferling, M.; Weiss, T.; Liu, N.; Giessen, H. *Nano Lett.* **2012**, *12*, 2542–2547.
(17) Yin, X.; Schäferling, M.; Metzger, B.; Giessen, H. *Nano Lett.* **2013**, *13*, 6238–6243.
(18) Cui, Y.; Kang, L.; Lan, S.; Rodrigues, S.; Cai, W. *Nano Lett.* **2014**, *14*, 1021–1025.
(19) Duan, X.; Yue, S.; Liu, N. *Nanoscale* **2015**, *7*, 17237–17243.
(20) Zhang, S.; Park, Y.-S.; Li, J.; Lu, X.; Zhang, W.; Zhang, X. *Phys. Rev. Lett.* **2009**, *102*, 023901.
(21) Zhang, S.; Zhou, J.; Park, Y.-S.; Rho, J.; Singh, R.; Nam, S.; Azad, A. K.; Chen, H.-T.; Yin, X.; Taylor, A. J.; Zhang, X. *Nat. Commun.* **2012**, *3*, 942.
(22) Cao, T.; Zhang, L.; Simpson, R. E.; Wei, C.; Cryan, M. J. *Opt. Express* **2013**, *21*, 27841–27851.
(23) Yin, X.; Schäferling, M.; Michel, A.-K. U.; Tittl, A.; Wuttig, M.; Taubner, T.; Giessen, H. *Nano Lett.* **2015**, *15*, 4255–4260.
(24) Borsa, D. M.; Baldi, A.; Pasturel, M.; Schreuders, H.; Dam, B.; Griessen, R.; Vermeulen, P.; Notten, P. H. L. *Appl. Phys. Lett.* **2006**, *88*, 241910.
(25) Baldi, A.; Palmisano, V.; Gonzalez-Silveira, M.; Pivak, Y.; Slaman, M.; Schreuders, H.; Dam, B.; Griessen, R. *Appl. Phys. Lett.* **2009**, *95*, 071903.
(26) Baldi, A.; Pálsson, G. K.; Gonzalez-Silveira, M.; Schreuders, H.; Slaman, M.; Rector, J. H.; Krishnan, G.; Kooi, B. J.; Walker, G. S.; Fay, M. W.; Hjörvarsson, B.; Wijngaarden, R. J.; Dam, B.; Griessen, R. *Phys. Rev. B* **2010**, *81*, 224203.
(27) Sanz, J. M.; Ortiz, D.; de la Osa, R. A.; Saiz, J. M.; González, F.; Brown, A. S.; Losurdo, M.; Everitt, H. O.; Moreno, F. *J. Phys. Chem. C* **2013**, *117*, 19606–19615.
(28) Ares, J. R.; Leardini, F.; Díaz-Chao, P.; Ferrer, I. J.; Fernández, J. F.; Sánchez, C. *Int. J. Hydrogen Energy* **2014**, *39*, 2587–2596.
(29) Sterl, F.; Strohfeldt, N.; Walter, R.; Griessen, R.; Tittl, A.; Giessen, H. *Nano Lett.* **2015**, *15*, 7949–7955.
(30) Strohfeldt, N.; Tittl, A.; Schäferling, M.; Neubrech, F.; Kreibig, U.; Griessen, R.; Giessen, H. *Nano Lett.* **2014**, *14*, 1140–1147.
(31) Liu, N.; Tang, M. L.; Hentschel, M.; Giessen, H.; Alivisatos, A. P. *Nat. Mater.* **2011**, *10*, 631–636.
(32) Tittl, A.; Mai, P.; Taubert, R.; Dregely, D.; Liu, N.; Giessen, H. *Nano Lett.* **2011**, *11*, 4366–4369.
(33) Tittl, A.; Yin, X.; Giessen, H.; Tian, X.-D.; Tian, Z.-Q.; Kremers, C.; Chigrin, D. N.; Liu, N. *Nano Lett.* **2013**, *13*, 1816–1821.
(34) Tittl, A.; Giessen, H.; Liu, N. *Nanophotonics* **2014**, *3*, 157–180.
(35) Griessen, R.; Strohfeldt, N.; Giessen, H. *Nat. Mater.* **2015**, DOI:10.1038/nmat4480.
(36) Nordlien, J.; Ono, S.; Masuko, N.; Nis, K. *J. Electrochem. Soc.* **1995**, *142*, 3320−3322.
(37) Lindström, R.; Johansson, L.-G.; Thompson, G. E.; Skeldon, P.; Svensson, J.-E. *Corros. Sci.* **2004**, *46*, 1141−1158.

TOC

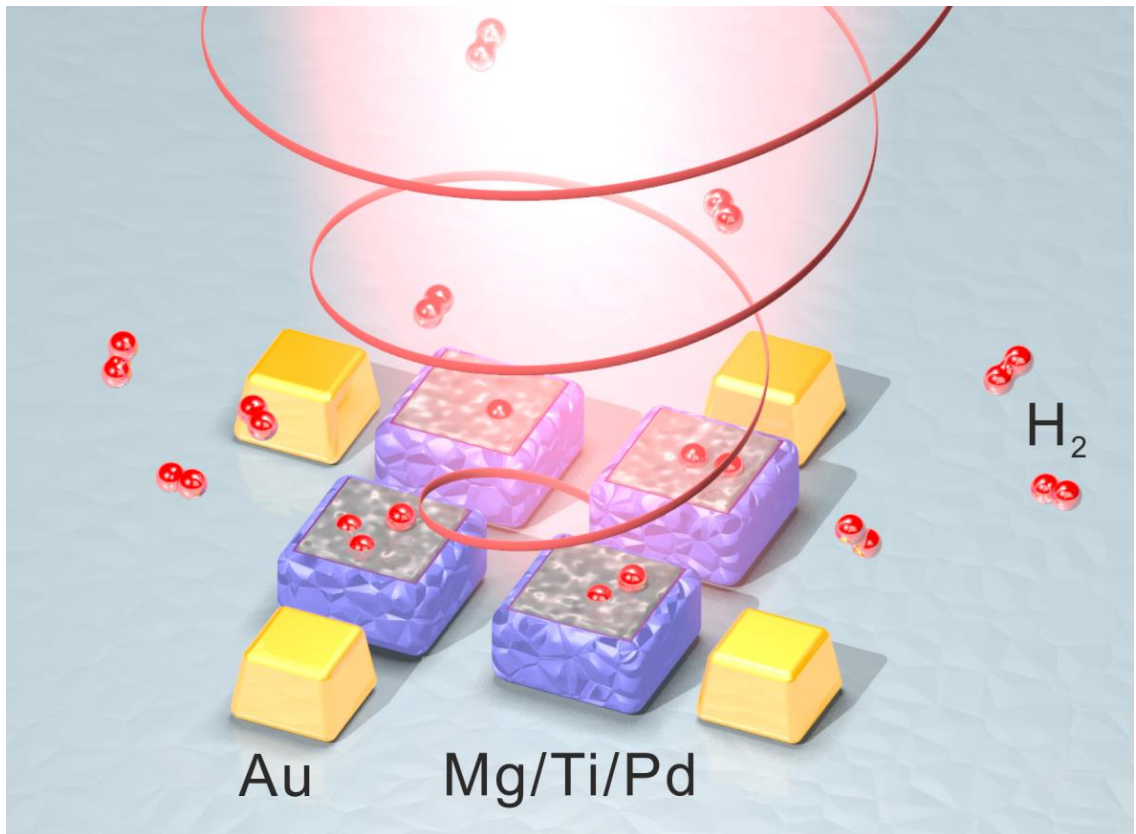